\documentclass[journal]{IEEEtran}
\pdfoutput=1
\usepackage{epsfig,latexsym}
\usepackage{float}
\usepackage{indentfirst}
\usepackage{amsmath}
\usepackage{amssymb}
\usepackage{times}
\usepackage{subfigure}
\usepackage{psfrag}
\usepackage{cite}
\usepackage{color}
\usepackage{algorithm}
\usepackage{setspace}
\usepackage{mathrsfs}
\usepackage{stfloats}
\usepackage{bm}
\makeatletter
\renewcommand{\maketag@@@}[1]{\hbox{\m@th\normalsize\normalfont#1}}%
\makeatother

\begin{document}

\title{\LARGE{Joint Optimization of STAR-RIS Assisted UAV Communication Systems}}

\author{Qin Zhang, Yang Zhao, Hai Li, Shujuan Hou, and Zhengyu Song

\thanks{Manuscript received June 29, 2022; revised August 31, 2022; accepted September 2, 2022. Date of publication xxx xx, xxxx. The associate editor coordinating the review of this article and approving it for publication was C.-K. Wen.
\emph{(Corresponding author: Hai Li.)}

Q. Zhang, Y. Zhao, H. Li, and S. Hou are with the School of Information and Electronics, Beijing Institute of Technology, Beijing 100081, China (e-mail: zqbit@bit.edu.cn; 3120210784@bit.edu.cn; haili@bit.edu.cn; shujuanhou@bit.edu.cn).

Z. Song is with the School of Electronic and Information Engineering, Beijing Jiaotong University, Beijing 100044, China (e-mail: songzy@bjtu.edu.cn).}
\vspace{-0.3in}
}

\markboth{IEEE WIRELESS COMMUNICATIONS LETTERS,~Vol.~xx, No.~xx, xx~2022}
{Shell \MakeLowercase{\textit{et al.}}: Bare Demo of IEEEtran.cls for Journals}

\maketitle

\begin{abstract}
In this letter, we study the simultaneously transmitting and reflecting reconfigurable intelligent surface (STAR-RIS) assisted unmanned aerial vehicle (UAV) communications. Our goal is to maximize the sum rate of all users by jointly optimizing the STAR-RIS’s beamforming vectors, the UAV’s trajectory and power allocation. We decompose the formulated non-convex problem into three subproblems and solve them alternately to obtain the solution. Simulations show that: 1) the STAR-RIS achieves a higher sum rate than traditional RIS; 2) to exploit the benefits of STAR-RIS, the UAV's trajectory is closer to STAR-RIS than that of RIS; 3) the energy splitting for reflection and transmission highly depends on the real-time trajectory of UAV.
\end{abstract}


\vspace{-0.12in}
\section{Introduction}
\vspace{-0.05in}
\IEEEPARstart{R}ECENTLY, the reconfigurable intelligent surface (RIS) has been considered as an emerging technology for future wireless communications \cite{wu2019towards}. However, the RIS is only able to reflect the incident signals, which means it can only achieve half-space coverage. To overcome this drawback, a novel concept of simultaneously transmitting and reflecting RIS (STAR-RIS) has been proposed \cite{niu2021simultaneous, niu2022weighted}. Different from the traditional RIS, the scattering elements of the STAR-RIS can reflect and transmit the incident signals at the same time, and thus achieve full-space coverage. More importantly, the STAR-RIS provides extra degrees-of-freedom (DoFs) by manipulating both the transmitting and reflecting signals \cite{liu2021star}.

Since the STAR-RIS offers more advantages than traditional RIS, there have been several studies on the STAR-RIS assisted communications \cite{liu2021star, zuo2021joint, mu2021simultaneously, niu2022weighted2, liu2021simultaneously}. In \cite{liu2021star}, three typical operating protocols for STAR-RIS were introduced, and the superiority of STAR-RIS over RIS was revealed by simulations. In \cite{zuo2021joint}, the sum rate maximization problem for STAR-RIS-NOMA systems was studied by jointly optimizing the decoding order, power allocation, and passive beamforming at the STAR-RIS. Focusing on the power consumption minimization problem, the beamforming of base station and STAR-RIS were jointly optimized for a STAR-RIS aided communication system in \cite{mu2021simultaneously}. Besides, in \cite{niu2022weighted2, liu2021simultaneously}, a more practical model with coupled transmitting and reflecting coefficients was considered to study the resource allocation in STAR-RIS-aided networks.

In recent years, the RIS assisted UAV communications have been intensively studied \cite{li2020reconfigurable, hua2021uav, li2021robust}, where the UAV's trajectory, RIS’s passive beamforming and/or active beamforming at the UAV were jointly optimized. Despite the benefits of STAR-RIS, there has been no work focusing on the interplay between STAR-RIS and UAV communications, where the coupled UAV's trajectory and STAR-RIS's transmission/reflection coefficient require careful design. In addition, the potential performance gain brought by STAR-RIS in UAV communications is still unknown. Motivated by these observations, in this letter, a novel STAR-RIS assisted UAV communication framework is proposed. We aim to maximize the sum rate of all users by jointly optimizing the STAR-RIS’s beamforming vectors, the UAV’s trajectory, and power allocation. To address the formulated non-convex problem, we decompose it into three subproblems and solve them alternately to obtain the solution. Simulations show that STAR-RIS can achieve higher sum rate than RIS. Besides, compared to the trajectory with the deployment of RIS, the UAV's trajectory with STAR-RIS is closer to the location of STAR-RIS in order to fully exploit the extra DoFs and acquire larger channel gains. More interestingly, the energy splitting for reflection and transmission modes highly depends on the real-time trajectory of UAV.

\vspace{-0.1in}
\section{System Model and Problem Formulation}
\subsection{System Model}
\begin{figure}[t]
\centering
\includegraphics[width = 3in]{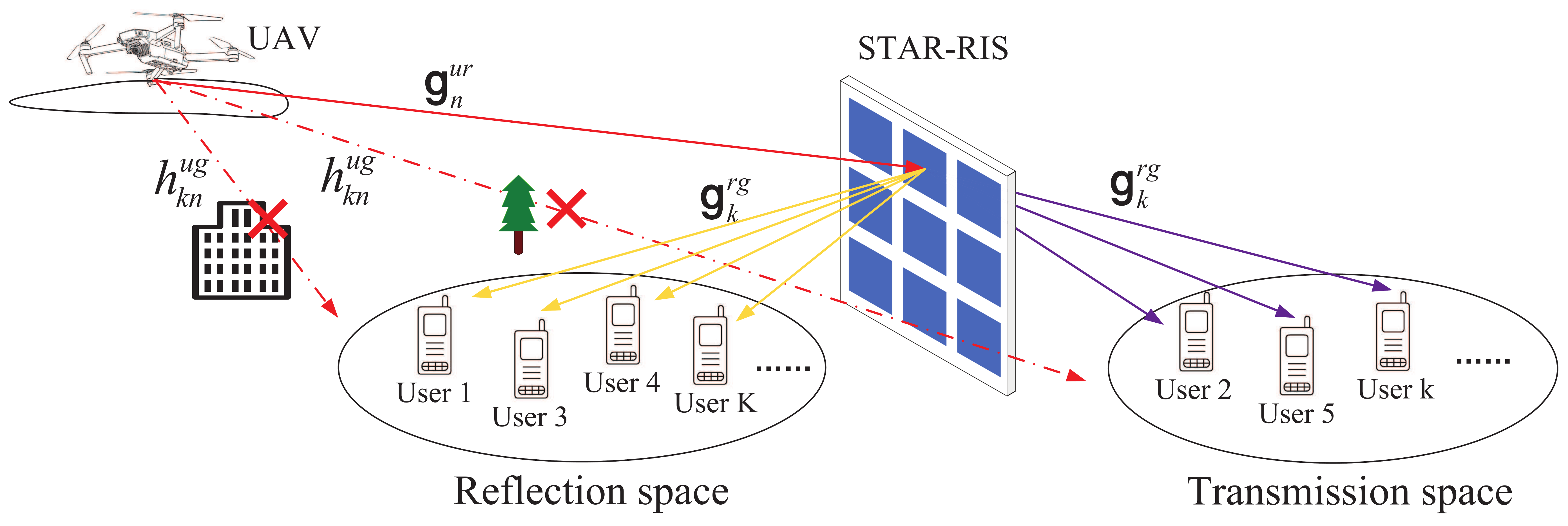}
\vspace{-0.1in}
\caption{Illustration of the STAR-RIS assisted UAV communication systems.}
\label{System}
\vspace{-0.15in}
\end{figure}

As illustrated in Fig. 1, we consider a STAR-RIS assisted UAV communication system, consisting of one rotary wing UAV, one STAR-RIS, and \(K\) ground users, where all users can access the UAV by NOMA. The UAV flies at a fixed altitude of \(H\). The flight time is decomposed into \(N\) equal time slots with sufficiently small slot length. Then the horizontal position of the UAV can be expressed by \(N + 1\) points in the two-dimensional plane, which is denoted by \({{\mathbf{q}}_n} = {\left[ {{x_n},{y_n}} \right]^T} \in {\mathbb{R}^{2 \times 1}}\), \(\forall n \in \left\{ {1, 2,\ldots ,N + 1} \right\}\). The mobility constraints of the UAV are \({{\bf{q}}_1} = {\bf{q}}_{\rm{1}}^{\rm{0}}, {{\bf{q}}_{{\rm{N + 1}}}} = {\bf{q}}_{{\rm{N + 1}}}^{\rm{0}}\) and \({\left\| {{{\bf{q}}_{n + 1}} - {{\bf{q}}_n}} \right\|^2} \le {D^2}\), where \({\mathbf{q}}_{\text{1}}^{\text{0}}\) and \({\mathbf{q}}_{{\text{N + 1}}}^{\text{0}}\) represent the initial and final locations of the UAV, and \(D\) represents the maximum distance that the UAV can move within a time slot.

We assume that the STAR-RIS is deployed at a fixed height of \({z_R}\), equipped with $M$ passive elements, and the horizontal position of the first element is denoted by \({{\bm{\omega }}_R} \!=\! {\left[ {{x_R},{y_R}} \right]^T} \in {\mathbb{R}^{2 \times 1}}\). There are $R$ and $T$ users in the reflection and transmission spaces of the STAR-RIS, respectively, and $R + T \!=\! K$. Also, the location of the $k$-th user is expressed as ${{\bm{\omega }}_k} \!=\! {\left[ {{x_k},{y_k}} \right]^T} \in {{\mathbb R}^{2 \times {\text{1}}}}$, $\forall k \! \in \! \left\{ {1,2, \ldots ,K} \right\}$. Consider the STAR-RIS works in the energy splitting (ES) mode, i.e., the energy of the signal incident on each element is split into the energies of transmitted and reflected signals \cite{mu2021simultaneously}. Let \({\mathbf{u}}_n^p = {\left[ {\sqrt {\beta _{1n}^p} {e^{j\theta _{1n}^p}},\sqrt {\beta _{2n}^p} {e^{j\theta _{2n}^p}}, \ldots ,\sqrt {\beta _{mn}^p} {e^{j\theta _{mn}^p}}, \ldots ,\sqrt {\beta _{Mn}^p} {e^{j\theta _{Mn}^p}}} \right]^H}\\ \in {\mathbb{C}^{M \times 1}}\) be the reflection ($p = r$) or transmission ($p = t$) beamforming vectors of the $m{\text{-th}}$ element at time slot $n$, and the corresponding diagonal matrix is denoted by \({\mathbf{\Phi }}_n^p = \rm{diag}({\mathbf{u}}_n^p) \in {\mathbb{C}^{M \times M}}\), where \(\left\{ {\sqrt {\beta _{mn}^r} ,\sqrt {\beta _{mn}^t}  \in [0,1]} \right\}\) and \(\left\{ {\theta _{mn}^r,\theta _{mn}^t \in [0,2\pi )} \right\}\) denote the amplitudes and phase shifts of the reflection and transmission coefficients for the $m{\text{-th}}$ element at time slot \(n\), respectively. Meanwhile, according to the energy conservation law, \(\beta _{mn}^r + \beta _{mn}^t = 1\), \(\forall m \in \{ 1, \ldots ,M\} \), \(\forall n \in \{ 1, \ldots ,N\} \), should be satisfied.

We assume the perfect channel state information (CSI) of all channels is available at the UAV\footnote{The CSI can be obtained by recently proposed channel estimation techniques for RIS-aided systems. With the perfect CSI assumption, the result provided in this letter is actually a theoretical performance upper bound.}, and the optimized transmission strategy can be exchanged among the STAR-RIS, UAV and ground users by a dedicated control channel. As shown in Fig. 1, since the STAR-RIS is generally deployed to assist transmissions where the line-of-sight (LoS) links are blocked by obstacles, the direct link from the UAV to each ground user is a Rayleigh fading channel \cite{li2020reconfigurable}. Therefore, the channel vector of the UAV to user $k$ at time slot $n$ is
\begin{equation}
h_{kn}^{ug} = \sqrt {\frac{\xi }{{{{(d_{kn}^{ug})}^{{o_1}}}}}} \tilde h_{kn}^{ug},
\end{equation}
where $\xi $ is the path loss at a reference distance of 1 meter, $d_{kn}^{ug} = \sqrt {{H^2} + ||{{\mathbf{q}}_n} - {{\bm{\omega }}_k}|{|^2}} $ denotes the distance between the UAV and the $k{\text{-th}}$ ground user at the $n{\text{-th}}$ time slot, ${o_1}$ represents the path loss exponent, and \(\tilde h_{kn}^{ug}\) is complex Gaussian random variable with zero mean and unit variance.

The transmission link from the UAV to STAR-RIS and from the STAR-RIS to ground users are assumed to be LoS channels. Therefore, at the $n{\text{-th}}$ time slot, the channel vector of the UAV-STAR-RIS link is given by
\begin{equation}
\bm{g}_n^{ur} = \sqrt {\frac{\xi }{{{{\left( {d_n^{ur}} \right)}^{{o_2}}}}}} {[1,{e^{ - j\frac{{2\pi }}{\lambda }d\varphi _n^{ur}}}, \ldots ,{e^{ - j\frac{{2\pi }}{\lambda }(M - 1)d\varphi _n^{ur}}}]^T},
\end{equation}
where $d_n^{ur} = \sqrt {{{(H - {z_R})}^2} + ||{{\mathbf{q}}_n} - {{\bm{\omega }}_R}|{|^2}} $ denotes the distance between the UAV and the STAR-RIS at the $n{\text{-th}}$ time slot, ${o_2}$ represents the corresponding path loss exponent, and $\varphi _n^{ur} = \frac{{{x_R} - {x_n}}}{{d_n^{ur}}}$ is the cosine of the angle of arrival (AoA) of the signal. And we define \({\bf{\Theta }}_n = {[1,{e^{ - j\frac{{2\pi }}{\lambda }d\varphi _n^{ur}}}, \ldots ,{e^{ - j\frac{{2\pi }}{\lambda }(M - 1)d\varphi _n^{ur}}}]^T}\).

Similarly, the channel vector of the STAR-RIS to the $k{\text{-th}}$ ground user is given by
\begin{equation}
\bm{g}_k^{rg} = \sqrt {\frac{\xi }{{{{\left( {d_k^{rg}} \right)}^{{o_3}}}}}} {[1,{e^{ - j\frac{{2\pi }}{\lambda }d\varphi _k^{rg}}},...,{e^{ - j\frac{{2\pi }}{\lambda }(M - 1)d\varphi _k^{rg}}}]^T},
\end{equation}
where $d_k^{rg} = \sqrt {{z_R}^2 + ||{{\bm{\omega }}_R} - {{\bm{\omega }}_k}|{|^2}} $ denotes the distance between the STAR-RIS and the $k{\text{-th}}$ ground user, ${o_{\text{3}}}$ represents the path loss exponent, and $\varphi _k^{rg} = \frac{{{x_R} - {x_k}}}{{d_k^{rg}}}$ is the cosine of the angle of departure (AoD) of the signal to the $k{\text{-th}}$ ground user. Therefore, at the $n{\text{-th}}$ time slot, the combined channel vector from the UAV to  ground user $k$ can be expressed as
\begin{equation}\label{g_kn}
{g_{kn}} = h_{kn}^{ug} + {(\bm{g}_k^{rg})^H}{\mathbf{\Phi }}_n^p\bm{g}_n^{ur}.
\end{equation}

For NOMA systems, the successive interference cancellation (SIC) is adopted by ground users to decode their own signals. Unfortunately, there are as many as \(K!\) decoding orders for \(K\) users, and the decoding order is highly coupled with STAR-RIS's beamforming vector design \cite{mu2020exploiting}. In order to reduce the complexity, we assume the decoding order has been predetermined, which is from user 1 to \(K\). As a result, the condition ${\left| {{g_{Kn}}} \right|^2} \geqslant {\left| {{g_{(K - 1)n}}} \right|^2} \geqslant  \ldots  \geqslant {\left| {{g_{1n}}} \right|^2}$ should be satisfied. Besides, ${p_{1n}} \geqslant  \ldots  \geqslant {p_{(K - 1)n}} \geqslant {p_{Kn}} \geqslant {\rm{0}}$ is used to ensure the fairness among users, where ${p_{kn}}$ denotes the transmission power allocated by the UAV to user $k$ at time slot $n$. Thus, at time slot $n$, the date rate of user $k$ is
\begin{equation}
{R_{kn}} = {\log _2}\left( {1 + \frac{{{p_{kn}}|{g_{kn}}{|^2}}}{{\sum\nolimits_{i = k + 1}^K {{p_{in}}|{g_{kn}}{|^2}}  + {\sigma ^2}}}} \right).
\end{equation}

\vspace{-0.1in}
\subsection{Problem Formulation}
Let ${\bf{Q}} = \left\{ {{{\mathbf{q}}_n},\forall n} \right\}$ denote the UAV's trajectory, ${\bf{U}} = \left\{ {{\mathbf{u}}_n^p,\forall n,\forall p \in \left\{ {t,r} \right\}} \right\}$ indicate the beamforming vectors of the STAR-RIS, and ${\bf{P}} = \left\{ {{p_{kn}},\forall n, k} \right\}$ represent the power allocation. In this paper, our goal is to maximize the sum rate of all users by jointly optimizing $\bf{Q}$, $\bf{U}$ and $\bf{P}$. Therefore, the optimization problem can be formulated as
\begin{subequations}\label{maxRsum}
\begin{align}
&\mathcal{P}:{\text{ }}\mathop {\max }\limits_{\bf{Q},\bf{U},\bf{P}} \sum\nolimits_{{n} = 1}^N {\sum\nolimits_{k = 1}^K {R_{kn}} }\\
&\rm{s.t.}\;{{\mathbf{q}}_1} = {\mathbf{q}}_{\text{1}}^{\text{0}}, {{\mathbf{q}}_{{\text{N + 1}}}} = {\mathbf{q}}_{{\text{N + 1}}}^{\text{0}},\label{R_max1Con}\\
&{\left\| {{{\mathbf{q}}_{n + 1}} - {{\mathbf{q}}_n}} \right\|^2} \leqslant {D^2}, \forall n,\label{R_max2Con}\\
&\beta _{mn}^r,\beta _{mn}^t \in [0,1], \forall m, \forall n,\label{R_max3Con}\\
&\theta _{mn}^r,\theta _{mn}^t \in \left[ {0,2\pi } \right), \forall m, \forall n,\label{R_max4Con}\\
&\beta _{mn}^r + \beta _{mn}^t = 1, \forall m, \forall n,\label{R_max5Con}\\
&\sum\nolimits_{k = 1}^K {{p_{kn}}}  \leqslant {P_{\max }}, \forall n, \label{R_max6Con}\\
&{\left| {{g_{Kn}}} \right|^2} \geqslant {\left| {{g_{(K - 1)n}}} \right|^2} \geqslant  \ldots  \geqslant {\left| {{g_{1n}}} \right|^2}, \forall n, \label{R_max7Con}\\
&{p_{1n}} \geqslant  \ldots  \geqslant {p_{(K - 1)n}} \geqslant {p_{Kn}} \geqslant {\text{0}}, \forall n \label{R_max8Con},
\end{align}
\end{subequations}
where ${P_{\max }}$ is the maximum transmit power of the UAV.

\vspace{-0.1in}
\section{Proposed Algorithm}
Problem $\mathcal{P}$ is difficult to solve because of the coupled variables and non-convex constraints. Therefore, we decompose $\mathcal{P}$ into three subproblems, where the STAR-RIS’s beamforming vectors, the UAV’s trajectory and power allocation are optimized, respectively.

\vspace{-0.1in}
\subsection{STAR-RIS's Beamforming Vectors Optimization}
When the UAV’s trajectory $\bf{Q}$ and the power allocation $\bf{P}$ are fixed, problem \(\mathcal{P}\) can be rewritten as
\begin{subequations}\label{maxRsum1}
\begin{align}
&\mathcal{P}1:{\text{ }}\mathop {\max }\limits_{\bf{U}} \sum\nolimits_{n = 1}^N {\sum\nolimits_{k = 1}^K {R_{kn}} }\\
&\rm{s.t.}\, \eqref{R_max3Con}, \eqref{R_max4Con}, \eqref{R_max5Con}, \eqref{R_max7Con}.
\end{align}
\end{subequations}

In order to solve $\mathcal{P}1$, we introduce the slack variables ${A_{kn}}$ and ${B_{kn}}$, where \(1/{A_{kn}} = {p_{kn}}{\left| {{g_{kn}}} \right|^2}\) and \({B_{kn}} = \sum\nolimits_{i = k + 1}^K {{p_{in}}{{\left| {{g_{kn}}} \right|}^2} + {\sigma ^2}}\). Thus, the achievable rate of user $k$ at time slot $n$ can be given by
\begin{equation}
{R_{kn}} = {\log _2}\left( {1 + \frac{1}{{{A_{kn}}{B_{kn}}}}} \right).
\end{equation}

Define ${{\mathbf{v}}_{kn}} = \left[ {{{\left( {\bm{g}_k^{rg}} \right)}^H}{\text{diag}}\left( {\bm{g}_n^{ur}} \right),h_{kn}^{ug}} \right]$, ${{\mathbf{V}}_{kn}} = {\mathbf{v}}_{kn}^H{{\mathbf{v}}_{kn}}$,  ${\mathbf{e}}_n^p = \left[ {{\mathbf{u}}_n^p;1} \right]$ and ${\mathbf{E}}_n^p = {\mathbf{e}}_n^p{\left( {{\mathbf{e}}_n^p} \right)^H}$, where \({\mathbf{E}}_n^p\underline  \succ  0\), \({\text{rank}}\left( {{\mathbf{E}}_n^p} \right) = 1\), \({\left[ {{\mathbf{E}}_n^p} \right]_{m,m}} = \beta _{mn}^p\), \({\left[ {{\mathbf{E}}_n^p} \right]_{M + 1,M + 1}} = 1\), \(p \in \{ r,t\} \), and ${\left[  \bullet  \right]_{m,m}}$ represents the $\left( {m,m} \right)$-th element value of the matrix. Therefore, according to \eqref{g_kn}, the combined channel gain from the UAV to user $k$ can be expressed as ${\left| {{g_{kn}}} \right|^2}{\rm{ = Tr}}\left( {{{\bf{V}}_{kn}}{\bf{E}}_n^p} \right)$.

Then, problem \(\mathcal{P}1\) can be transformed as
\begin{subequations}
\begin{align}
&\mathcal{P}2:{\text{ }}\mathop {\max }\limits_{{\mathbf{E}}_n^p,{A_{kn}},{B_{kn}},{R_{kn}}} \sum\nolimits_{n = 1}^N {\sum\nolimits_{k = 1}^K {{R_{kn}}} } \\
&\rm{s.t.}\;\eqref{R_max5Con},\\
&{\log _2}\left( {1 + \frac{1}{{{A_{kn}}{B_{kn}}}}} \right) \geqslant {R_{kn}},\label{R_max2_1Con}\\
&\frac{1}{{{A_{kn}}}} \leqslant {\text{Tr}}\left( {{{\mathbf{V}}_{kn}}{\mathbf{E}}_n^p} \right){p_{kn}},\label{R_max2_2Con}\\
&{B_{kn}} \geqslant {\text{Tr}}\left( {{{\mathbf{V}}_{kn}}{\mathbf{E}}_n^p} \right)\sum\nolimits_{i = k + 1}^K {{p_{in}} + {\sigma ^2}},\\
&{\mathbf{E}}_n^p\underline  \succ  0\ ,\\
&{\left[ {{\mathbf{E}}_n^p} \right]_{m,m}} = \beta _{mn}^p,\\
&{\left[ {{\mathbf{E}}_n^p} \right]_{M + 1,M + 1}} = 1,\label{R_max2_6Con}\\
&{\text{rank}}\left( {{\mathbf{E}}_n^p} \right) = 1,\label{R_max2_7Con}\\
&{\rm{Tr}}\left( {{{\bf{V}}_{Kn}}{\bf{E}}_n^p} \right) \geqslant {\rm{Tr}}\left( {{{\bf{V}}_{(K\!-\! 1)n}}{\bf{E}}_n^p} \right) \geqslant \!\ldots\!  \geqslant {\rm{Tr}}\left( {{{\bf{V}}_{1n}}{\bf{E}}_n^p} \right)\label{R_max2_8Con}.
\end{align}
\end{subequations}

For non-convex constraint \eqref{R_max2_1Con}, we apply the first-order Taylor expansion at local points \(A_{kn}^{({\tau _1})}\) and \(B_{kn}^{({\tau _1})}\) obtained in the ${\tau _1}$-th iteration. Then, constraint \eqref{R_max2_1Con} is approximated by
\begin{equation}
\begin{array}{l}
\!{\log _2}\!\left(\!{1\!+\!\frac{1}{{A_{kn}^{({\tau _1})}B_{kn}^{({\tau _1})}}}}\!\right)\!-\!\frac{{{{\log }_2}e\left(\!{{A_{kn}}\!-\!A_{kn}^{({\tau _1})}}\!\right)}}{{A_{kn}^{({\tau _1})}\left(\!{1\!+\!A_{kn}^{({\tau _1})}B_{kn}^{({\tau _1})}}\!\right)}}\!-\!\frac{{{{\log }_2}e\left(\! {{B_{kn}}\!-\!B_{kn}^{({\tau _1})}}\!\right)}}{{B_{kn}^{({\tau _1})}\left(\!{1\!+\!A_{kn}^{({\tau _1})}B_{kn}^{({\tau _1})}}\!\right)}}\\
\! \geqslant {R_{kn}}.\label{R_max2_3Con}
\end{array}
\end{equation}

However, the optimization problem is still non-convex due to constraint \eqref{R_max2_7Con}. Therefore, we propose the sequential rank-one constraint relaxation (SROCR) approach to obtain a rank-one solution. Firstly, the non-convex rank-one constraint \eqref{R_max2_7Con} can be relaxed as \cite{mu2020exploiting}
\begin{equation}
{\varepsilon _{\max }}\left( {{\mathbf{E}}_n^p} \right) \geqslant {\alpha ^{({\tau _2})}}{\text{Tr}}\left( {{\mathbf{E}}_n^p} \right),
\end{equation}
where \({\varepsilon _{\max }}\left( {{\mathbf{E}}_n^p} \right)\) is the largest eigenvalue of \({\mathbf{E}}_n^p\), and \({\alpha ^{({\tau _2})}} \in \left[ {0,1} \right]\) is a relaxation parameter in the ${\tau _2}{\text{-th}}$ iteration. And $\alpha ^{({\tau _2})}$ can be updated with the following formula \cite{mu2020exploiting}
\begin{equation}\label{UPalpha}
{\alpha ^{({\tau _2})}} = \min \left( {1,\frac{{{\varepsilon _{\max }}\left( {{\mathbf{E}}{{_n^p}^{({\tau _2})}}} \right)}}{{{\text{Tr}}\left( {{\mathbf{E}}{{_n^p}^{({\tau _2})}}} \right)}} + {\Delta ^{({\tau _2})}}} \right),
\end{equation}
where \({\Delta ^{({\tau _2})}}\) denotes the step size in the ${\tau _2}{\text{-th}}$ iteration and needs to be resized according to the problem. Furthermore, since \({\varepsilon _{\max }}\left( {{\mathbf{E}}_n^p} \right)\) is non-differentiable, it can be approximated by the expression \({\varepsilon _{\max }}\left( {{\mathbf{E}}_n^p} \right) = {\mathbf{e}}_{\max }^H\left( {{\mathbf{E}}{{_n^p}^{({\tau _2})}}} \right){\mathbf{E}}_n^p{{\mathbf{e}}_{\max }}\left( {{\mathbf{E}}{{_n^p}^{({\tau _2})}}} \right)\), where \({{\mathbf{e}}_{\max }}\left( {{\mathbf{E}}{{_n^p}^{({\tau _2})}}} \right)\) denotes the eigenvector corresponding to the largest eigenvalue of \({{\mathbf{E}}{{_n^p}^{({\tau _2})}}}\), and \({{\mathbf{E}}{{_n^p}^{({\tau _2})}}}\) is the solution in the ${\tau _2}{\text{-th}}$ iteration.

Finally, the STAR-RIS's beamforming vectors optimization problem can be approximated as
\begin{subequations}\label{maxRsum3}
\begin{align}
&\mathcal{P}3:{\text{ }}\mathop {\max }\limits_{{\mathbf{E}}_n^p,{A_{kn}},{B_{kn}},{R_{kn}}} \sum\nolimits_{n = 1}^N {\sum\nolimits_{k = 1}^K {{R_{kn}}} } \\
&\rm{s.t.}\, \eqref{R_max5Con}, \eqref{R_max2_2Con}-\eqref{R_max2_6Con}, \eqref{R_max2_8Con},\eqref{R_max2_3Con},\\
&e_{\max }^H\left( {{\mathbf{E}}{{_n^p}^{({\tau _2})}}} \right){\mathbf{E}}_n^p{e_{\max }}\left( {{\mathbf{E}}{{_n^p}^{({\tau _2})}}} \right) \geqslant {\alpha ^{({\tau _2})}}{\text{Tr}}\left( {{\mathbf{E}}_n^p} \right).
\end{align}
\end{subequations}

Problem $\mathcal{P}3$ without the rank-one constraint \eqref{R_max2_7Con} is a standard convex semidefinite programming (SDP) problem \cite{zuo2021joint}, which can be effectively solved by standard solvers. Thus, the algorithm for solving \(\mathcal{P}1\) can be summarized as Algorithm 1.

\begin{algorithm}[t]
\small
\caption{Algorithm for STAR-RIS's Beamforming Vectors Optimization}
\centering
\begin{tabular}{p{8cm}}
\noindent\hangafter=1\setlength{\hangindent}{1.2em}1. Initialize \({{\bf{U}}^{(0)}}\), \({\Delta ^{(0)}}\) and the error tolerance $\upsilon $, and calculate \({\mathbf{E}}_n^{p(0)}\). Set \({\alpha ^{(0)}} = 0\) and the iterative index ${\tau _2} = 0$.

\noindent\hangafter=1\setlength{\hangindent}{1.2em}2. \textbf{repeat}

\noindent\hangafter=1\setlength{\hangindent}{1.2em}3. \hspace{1em} Solve the problem \eqref{maxRsum3} to optimize \({\mathbf{E}}_n^p\);

\noindent\hangafter=1\setlength{\hangindent}{1.2em}4. \hspace{1em} \textbf{if} problem \eqref{maxRsum3} is solvable then

\noindent\hangafter=1\setlength{\hangindent}{1.2em}5. \hspace{2em} Update \({\mathbf{E}}_n^{p\left( {{\tau _2} + 1} \right)} = {\mathbf{E}}_n^p\), \({\Delta ^{({\tau _2} + 1)}} = {\Delta ^{({\tau _2})}}\);

\noindent\hangafter=1\setlength{\hangindent}{1.2em}6. \hspace{1em} \textbf{else}

\noindent\hangafter=1\setlength{\hangindent}{1.2em}7. \hspace{2em} Update \({\mathbf{E}}_n^{p\left( {{\tau _2} + 1} \right)} = {\mathbf{E}}_n^{p\left( {{\tau _2}} \right)}\), \({\Delta ^{({\tau _2} + 1)}} = \frac{{{\Delta ^{({\tau _2})}}}}{2}\);

\noindent\hangafter=1\setlength{\hangindent}{1.2em}8. \hspace{1em} \textbf{end}

\noindent\hangafter=1\setlength{\hangindent}{1.2em}9. \hspace{1em} Update \({\alpha ^{({\tau _2})}}\) by \eqref{UPalpha}, ${\tau _2} = {\tau _2} + 1$;

\noindent\hangafter=1\setlength{\hangindent}{1.2em}10. \textbf{until} \(|1 - {\alpha ^{({\tau _2})}}| \leqslant \upsilon \) and the objective of \eqref{maxRsum3} converges.

\noindent\hangafter=1\setlength{\hangindent}{1.2em}11. Obtain \({\mathbf{E}}{_n^{p^*}}\), and return the optimal solution \({{\bf{U}}^*}\).
\end{tabular}
\end{algorithm}

\begin{algorithm}[t]
\small
\caption{Algorithm for UAV’s Trajectory Optimization}
\centering
\begin{tabular}{p{8cm}}
\noindent\hangafter=1\setlength{\hangindent}{1.2em}1. Initialize \({{\bf{Q}}^{(0)}}\). Set the iterative index ${\tau _3} = 0$ and the threshold value $\delta $.

\noindent\hangafter=1\setlength{\hangindent}{1.2em}2. \textbf{repeat}

\noindent\hangafter=1\setlength{\hangindent}{1.2em}3. \hspace{1em} Solve the problem \eqref{maxRsum6} to optimize \(\bf{Q}\);

\noindent\hangafter=1\setlength{\hangindent}{1.2em}4. \hspace{1em} Update \({{\bf{Q}}^{({\tau _3} + 1)}} = \bf{Q}\), ${\tau _3} = {\tau _3} + 1$;

\noindent\hangafter=1\setlength{\hangindent}{1.2em}5. \textbf{until} the increase of the objective value is below $\delta $.

\noindent\hangafter=1\setlength{\hangindent}{1.2em}6. Return the optimal solution \({{\bf{Q}}^*}\).
\end{tabular}
\end{algorithm}

\vspace{-0.15in}
\subsection{UAV’s Trajectory Optimization}
If the STAR-RIS's beamforming vectors {\bf{U}} and the power allocation {\bf{P}} are given, problem \(\mathcal{P}\) can be reformulated as
\begin{subequations}\label{maxRsum4}
\begin{align}
&\mathcal{P}4:{\text{ }}\mathop {\max }\limits_{\bf{Q}} \sum\nolimits_{n = 1}^N {\sum\nolimits_{k = 1}^K {R_{kn}} } \\
&\rm{s.t.}\, \eqref{R_max1Con}, \eqref{R_max2Con}, \eqref{R_max7Con}.
\end{align}
\end{subequations}

To maximize the sum rate of users, it is assumed that the phase alignment of the signals at the users can be achieved \cite{li2020reconfigurable}. As such, ${\left| {{g_{kn}}} \right|^2}$ can be expressed as
\begin{equation}
\begin{gathered}
  {\left| {{g_{kn}}} \right|^2} = \frac{{{C_{kn}}}}{{{{\left( {d_n^{ur}} \right)}^{{o_2}}}}} + \frac{{{F_{kn}}}}{{{{\left( {d_n^{ur}} \right)}^{{o_2}/2}}{{\left( {d_{kn}^{ug}} \right)}^{{o_1}/2}}}} + \frac{{{G_{kn}}}}{{{{\left( {d_{kn}^{ug}} \right)}^{{o_1}}}}}, \hfill \\
\end{gathered}
\end{equation}
where \({C_{kn}}\!=\!\frac{{{\xi ^2}}}{{{{\left( {d_k^{rg}} \right)}^{{o_3}}}}}{\left| {{{\bf{\Theta }}_n^H}{{\left[ {\sqrt {\beta _{1n}^p} ,\sqrt {\beta _{2n}^p} , \ldots ,\sqrt {\beta _{Mn}^p} } \right]}^T}} \right|^2}\),\vspace{1ex} \\
 \({F_{kn}}\!\!=\!\!\frac{{2{\xi ^{3/2}}}}{{{{\left( {d_k^{rg}} \right)}^{{o_3}/2}}}}{\mathop{\rm Re}\nolimits}\! \!\left(\!{{{\bf{\Theta }}_n^H\!}{{\left( {\tilde h_{kn}^{ug}} \!\right)\!}^H}{{\!\left[\! {\sqrt {\beta _{1n}^p} ,\sqrt {\beta _{2n}^p} , \ldots ,\sqrt {\beta _{Mn}^p} } \right]\!}^T}} \right)\) and \({G_{kn}}\!=\!\xi {\left| {\tilde h_{kn}^{ug}} \right|^2}\).

\setlength{\lineskip}{2.5pt}
\setlength{\lineskiplimit}{2.5pt}
Then, slack variables ${j_{kn}} \geqslant d_{kn}^{ug} = \sqrt {{H^2} + ||{{\mathbf{q}}_n} - {{\bm{\omega }}_k}|{|^2}} $ and ${l_n} \geqslant d_n^{ur} = \sqrt {{{(H - {z_R})}^2} + ||{{\mathbf{q}}_n} - {{\bm{\omega }}_R}|{|^2}} $ are introduced, and problem \(\mathcal{P}4\) can be reformulated as
\setlength{\lineskip}{2.5pt}
\setlength{\lineskiplimit}{2.5pt}
\begin{subequations}\label{maxRsum5}
\begin{align}
&\mathcal{P}5:\mathop {\max }\limits_{{\bf{Q}},{l_n},{j_{kn}}} \sum\nolimits_{n = 1}^N {\sum\nolimits_{k = 1}^K {{{\log }_2}\left( {1\!+\! \frac{{{p_{kn}}}}{{\sum\nolimits_{i\!=\!k\!+\! 1}^K {{p_{in}}} \!+\!{\sigma ^2}/{S_{kn}}}}} \right)} } \label{maxRsum5_obj}\\
&\rm{s.t.}\, \eqref{R_max1Con}, \eqref{R_max2Con}, \\
&{l_n} \geqslant \sqrt {{{\left( {H - {z_R}} \right)}^2} + {{\left\| {{{\mathbf{q}}_n} - {{\bm{\omega }}_R}} \right\|}^2}},\\
&{j_{kn}} \geqslant \sqrt {{H^2} + {{\left\| {{{\mathbf{q}}_n} - {{\bm{\omega }}_k}} \right\|}^2}},\\
&{S_{Kn}} \geqslant {S_{(K - 1)n}} \geqslant  \ldots  \geqslant {S_{1n}},
\end{align}
\end{subequations}
where ${S_{kn}} = \frac{{{C_{kn}}}}{{{{\left( {{l_n}} \right)}^{{o_2}}}}} + \frac{{{F_{kn}}}}{{{{\left( {{l_n}} \right)}^{{o_2}/2}}{{\left( {{j_{kn}}} \right)}^{{o_1}/2}}}} + \frac{{{G_{kn}}}}{{{{\left( {{j_{kn}}} \right)}^{{o_1}}}}}$.

However, problem $\mathcal{P}5$ is still non-convex. Therefore, at the given points \({l_n^{({\tau _3})}}\) and \({j_{kn}^{({\tau _3})}}\), we derive the first-order Taylor expansions of \eqref{maxRsum5_obj}, \(S_{kn}\), \(l_n^2\) and \(j_{kn}^2\), respectively, and obtain
\vspace{-0.09in}
\begin{equation}
\begin{array}{l}
{\log _2}\left( {1\!+\!\frac{{{p_{kn}}}}{{\sum\nolimits_{i\!=\!k\!+\!1}^K {{p_{in}}} \!+\!{\sigma ^2}/{S_{kn}}}}} \right) \!\geqslant\! {\log _2}\left( {1\!+\!\frac{{{p_{kn}}}}{{\sum\nolimits_{i\!=\!k\!+\!1}^K {{p_{in}}}\! +\!{\sigma ^2}/S_{kn}^{({\tau _3})}}}} \right) \vspace{1ex}\\
 - W_{kn}^{\left( {{\tau _3}} \right)}L_{kn}^{\left( {{\tau _3}} \right)}\left( {{l_n} - l_n^{({\tau _3})}} \right) - W_{kn}^{\left( {{\tau _3}} \right)}J_{kn}^{\left( {{\tau _3}} \right)}\left( {{j_{kn}} - j_{kn}^{({\tau _3})}} \right),
\end{array}
\end{equation}
\begin{equation}
\begin{array}{l}
{S_{kn}}{\rm{ }} \geqslant {\rm{ }}\frac{{{C_{kn}}}}{{{{\left( {l_n^{({\tau _3})}} \right)}^{{o_2}}}}} + \frac{{{F_{kn}}}}{{{{\left( {l_n^{({\tau _3})}} \right)}^{{o_2}/2}}{{\left( {j_{kn}^{({\tau _3})}} \right)}^{{o_1}/2}}}} + \frac{G}{{{{\left( {j_{kn}^{({\tau _3})}} \right)}^{{o_1}}}}} \vspace{1ex}\\
\;\;\; - L_{kn}^{\left( {{\tau _3}} \right)}\left( {{l_n} - l_n^{({\tau _3})}} \right) - J_{kn}^{\left( {{\tau _3}} \right)}\left( {{j_{kn}} - j_{kn}^{({\tau _3})}} \right) = {{\rm O}_{kn}},
\end{array}
\end{equation}
\begin{equation}
l_n^2 \geqslant 2l_n^{({\tau _3})}l_n \!-\! {\left( {l_n^{({\tau _3})}} \right)^2}, \;\;\;j_{kn}^2 \geqslant 2j_{kn}^{({\tau _3})}j_{kn} \!-\! {\left( {j_{kn}^{({\tau _3})}} \right)^2},
\end{equation}
where \(W_{kn}^{\left( {{\tau _3}} \right)}\!=\!\frac{{{p_{kn}}{\sigma ^2}/{\rm{ln2}}}}{{\left( {\sum\nolimits_{i\!=\!k\!+\!1}^K {{p_{in}}}  + {\sigma ^2}/S_{kn}^{({\tau _3})}} \right)\left( {\sum\nolimits_{i\!=\!k}^K {{p_{in}}}  + {\sigma ^2}/S_{kn}^{({\tau _3})}} \right){{\left( {S_{kn}^{({\tau _3})}} \right)}^2}}}\), \(L_{kn}^{\left( {{\tau _3}} \right)}\!=\! \frac{{{o_2}{C_{kn}}}}{{{{\left( {l_n^{({\tau _3})}} \right)}^{{o_2}\!+\!1}}}}\!+\! \frac{{{o_2}{F_{kn}}/2}}{{{{\left( {l_n^{({\tau _3})}} \right)}^{{o_2}\!/\!2\!+\!1}}{{\left( {j_{kn}^{({\tau _3})}} \right)}^{{o_1}\!/\!2}}}}\), \(J_{kn}^{\left( {{\tau _3}} \right)}\!=\!\frac{{{o_1}G}}{{{{\left( {j_{kn}^{({\tau _3})}} \right)}^{{o_1}\!+\!1}}}}\) \\ \(+\frac{{{o_1}{F_{kn}}/2}}{{{{\left( {l_n^{({\tau _3})}} \right)}^{{o_2}}}{{\left( {j_{kn}^{({\tau _3})}} \right)}^{{o_1}\!/\!2\!+\!1}}}}\), and ${\tau _3}$ denotes the iterative index.

Finally, the UAV’s trajectory optimization problem can be transformed as
\begin{subequations}\label{maxRsum6}
\begin{align}
&\mathcal{P}6:{\text{ }}\mathop {\max }\limits_{{\bf{Q}},{l_n},{j_{kn}}} \sum\nolimits_{n = 1}^N {\sum\nolimits_{k = 1}^K {\left( { - {L_{kn}^{\left( {{\tau _3}} \right)}}{l_n} - {J_{kn}^{\left( {{\tau _3}} \right)}}{j_{kn}}} \right)} } \\
&\rm{s.t.}\, \eqref{R_max1Con}, \eqref{R_max2Con},\\
&{O_{Kn}} \geqslant {O_{(K\!-\!1)n}} \geqslant  \ldots  \geqslant {O_{1n}},\\
&{\left( {H\!-\!{z_R}} \right)^2}\!+\!{\left\| {{{\mathbf{q}}_n}\!-\!{{\bm{\omega }}_R}} \right\|^2}\!+\!{\left( {l_n^{({\tau _3})}} \right)^2}\!-\!2l_n^{({\tau _3})}l_n^{} \leqslant 0,\\
&{H^2}\!+\!{\left\| {{{\mathbf{q}}_n}\!-\! {{\bm{\omega }}_k}} \right\|^2}\!+\! {\left( {j_{kn}^{({\tau _3})}} \right)^2}\!-\! 2j_{kn}^{({\tau _3})}j_{kn}^{} \leqslant 0.
\end{align}
\end{subequations}

Similarly, problem $\mathcal{P}6$ is convex and can be solved efficiently by standard solvers such as CVX \cite{li2020reconfigurable}. The proposed algorithm for solving problem $\mathcal{P}4$ is shown in Algorithm 2.

\vspace{-0.1in}
\subsection{Power Allocation Optimization}
For the given UAV’s trajectory $\bf{Q}$ and STAR-RIS's beamforming vectors $\bf{U}$, problem \(\mathcal{P}\) can be reformulated as
\begin{subequations}\label{maxRsum7}
\begin{align}
&\mathcal{P}7:{\text{ }}\mathop {\max }\limits_{\bf{P}} \sum\nolimits_{n = 1}^N {\sum\nolimits_{k = 1}^K {R_{kn}}} \\
&\rm{s.t.}\, \eqref{R_max6Con}, \eqref{R_max8Con}.
\end{align}
\end{subequations}

\begin{algorithm}[t]
\small
\caption{Algorithm for Power Allocation Optimization}
\centering
\begin{tabular}{p{8cm}}
\noindent\hangafter=1\setlength{\hangindent}{1.2em}1. Initialize \({{\bf{P}}^{(0)}}\), and calculate $\left\{ {\psi _{kn}^{(0)}} \right\}$. Set the iterative index ${\tau _4} = 0$ and the threshold value $\delta '$.

\noindent\hangafter=1\setlength{\hangindent}{1.2em}2. \textbf{repeat}

\noindent\hangafter=1\setlength{\hangindent}{1.2em}3. \hspace{1em} Solve the problem \eqref{maxRsum9} to optimize $\left\{ {{\psi _{kn}}} \right\}$;

\noindent\hangafter=1\setlength{\hangindent}{1.2em}4. \hspace{1em} Update \(\left\{ {\psi _{kn}^{\left( {{\tau _3} + 1} \right)}} \right\} = \left\{ {{\psi _{kn}}} \right\}\), ${\tau _4} = {\tau _4} + 1$;

\noindent\hangafter=1\setlength{\hangindent}{1.2em}5. \textbf{until} the increase of the objective value is below $\delta '$.

\noindent\hangafter=1\setlength{\hangindent}{1.2em}6. Obtain $\left\{ {\psi _{kn}^*} \right\}$, and return the optimal solution \({{\bf{P}}^*}\).
\end{tabular}
\end{algorithm}

Let ${\psi _{kn}} = \sum\nolimits_{i = k}^K {{p_{in}}} $ and ${\psi _{(K + 1)n}} \triangleq 0$. Considering the first-order Taylor expansion of ${{\rm R}_{kn}}$ at the given point $\psi _{(k + 1)n}^{({\tau _4})}$, we have
\begin{equation}
\begin{array}{l}
{{\rm R}_{kn}} \geqslant {\overline {\rm R} _{kn}}\vspace{1ex}\\
 = {\log _2}\left( {{{\left| {{g_{kn}}} \right|}^2}{\psi _{kn}} + {\sigma ^2}} \right) - {\log _2}\left( {{{\left| {{g_{kn}}} \right|}^2}\psi _{(k + 1)n}^{({\tau _4})} + {\sigma ^2}} \right)\vspace{1ex}
\end{array}
\end{equation}

\vspace{-0.18in}
\({\rm{   }} - \frac{{{{\log }_2}e{{\left| {{g_{kn}}} \right|}^2}}}{{{{\left| {{g_{kn}}} \right|}^2}\psi _{(k + 1)n}^{({\tau _4})} + {\sigma ^2}}}\left( {{\psi _{(k + 1)n}} - \psi _{(k + 1)n}^{({\tau _4})}} \right)\),\vspace{0.1in}\\
where ${\tau _4}$ denotes the iterative index. Therefore, problem \(\mathcal{P}7\) can be approximated as
\begin{subequations}\label{maxRsum9}
\begin{align}
&\mathcal{P}8:{\text{ }}\mathop {\max }\limits_{\{ {\psi _{kn}}\} } \sum\nolimits_{n = 1}^N {\sum\nolimits_{k = 1}^K {{{\overline {\rm R} }_{kn}}} } \\
&{\rm{s.t.}}\;{\psi _{1n}} \leqslant {P_{\max }}, \label{R_Con1}\\
&{\psi _{1n}}\!-\!{\psi _{2n}} \geqslant  \ldots  \geqslant {\psi _{(K\!-\!1)n}}\!-\!{\psi _{Kn}} \geqslant {\psi _{Kn}} \geqslant 0. \label{R_Con2}
\end{align}
\end{subequations}

It is easy to prove that the problem $\mathcal{P}8$ is convex since it has a linear objective function and linear inequality constraints. After solving problem $\mathcal{P}8$, the power allocation can be obtained by $p_{kn}^* = \psi _{kn}^* - \psi _{(k + 1)n}^*$, $\forall k, n$. Then, the algorithm for solving problem $\mathcal{P}7$ is summarized in Algorithm 3.

\begin{algorithm}[t]
\small
\caption{Proposed Algorithm for Problem \({\cal P}\)}
\centering
\begin{tabular}{p{8cm}}
\noindent\hangafter=1\setlength{\hangindent}{1.2em}1. Initialize \({{\bf{U}}^{({\rm{0}})}}\), \({{\bf{Q}}^{({\rm{0}})}}\) and \({{\bf{P}}^{({\rm{0}})}}\). Set the iterative index ${\tau} = 0$ and the threshold value $\delta ''$.

\noindent\hangafter=1\setlength{\hangindent}{1.2em}2. \textbf{repeat}

\noindent\hangafter=1\setlength{\hangindent}{2.45em}3. \hspace{1em} Update \({{\bf{U}}^{(\tau  + 1)}}\) via \textbf{Algorithm 1} with \({{\bf{Q}}^{(\tau )}}\) and \({{\bf{P}}^{(\tau)}}\);

\noindent\hangafter=1\setlength{\hangindent}{2.45em}4. \hspace{1em} Update \({{\bf{Q}}^{(\tau  + 1)}}\) via \textbf{Algorithm 2} with \({{\bf{U}}^{(\tau  + 1)}}\) and \({{\bf{P}}^{(\tau)}}\);

\noindent\hangafter=1\setlength{\hangindent}{2.45em}5. \hspace{1em} Update \({{\bf{P}}^{(\tau + 1)}}\) via \textbf{Algorithm 3} with \({{\bf{U}}^{(\tau  + 1)}}\) and \({{\bf{Q}}^{(\tau  + 1)}}\) and $\tau  = \tau  + 1$;

\noindent\hangafter=1\setlength{\hangindent}{1.2em}6. \textbf{until} the increase of the objective value is below $\delta ''$.

\noindent\hangafter=1\setlength{\hangindent}{1.2em}7. Return the optimal solution \({{\bf{U}}^*}\), \({{\bf{Q}}^*}\) and \({{\bf{P}}^*}\).
\end{tabular}
\end{algorithm}

\vspace{-0.1in}
\subsection{Joint Beamforming Vectors, Power Allocation, and Trajectory Optimization Algorithm}
With the aid of Algorithms 1-3, the proposed algorithm to jointly optimize the STAR-RIS’s beamforming vectors, the UAV’s trajectory and power allocation is summarized in Algorithm 4. Since the objective value of problem $\mathcal{P}$ is non-decreasing after each iteration, the proposed algorithm is guaranteed to converge \cite{niu2022weighted}.

\vspace{-0.1in}
\section{Simulation Results}
\vspace{-0.03in}
In the simulations, we consider one STAR-RIS deployed at the height of \({z_R} = 20\) m, and the first element is located at \({{\bf{\omega }}_R} = {\left[ {{\rm{0}},{\rm{0}}} \right]^T}\)m in a horizontal plane. For a fair comparison between STAR-RIS and RIS, one transmitting-only RIS and one reflecting-only RIS with M/2 elements are employed at the same location as STAR-RIS. Besides, the mode selection (MS) protocol for STAR-RIS is also considered for comparison \cite{liu2021star}. The UAV flies at a fixed altitude $H \!=\! 30$ m, and its initial and final locations are \({\bf{q}}_{\rm{1}}^{\rm{0}} \!=\! {\left[ {10,-250} \right]^T}\)m and \({\bf{q}}_{{\rm{N + 1}}}^{\rm{0}} = {\left[ {10,250} \right]^T}\)m, respectively. The flight time of the UAV is decomposed into \(N = {\rm{30}}\) time slots, and ${P_{\max }} = 30$ dBm. The path loss exponents are ${o_1}{\text{ = 3}}$, ${o_2}{\text{ = 2}}$, and ${o_{\text{3}}}{\text{ = 2}}{\text{.8}}$ \cite{li2020reconfigurable}.

\begin{figure*}[!htp]
\centering
\subfigure[]{
\includegraphics[width=2.25in]{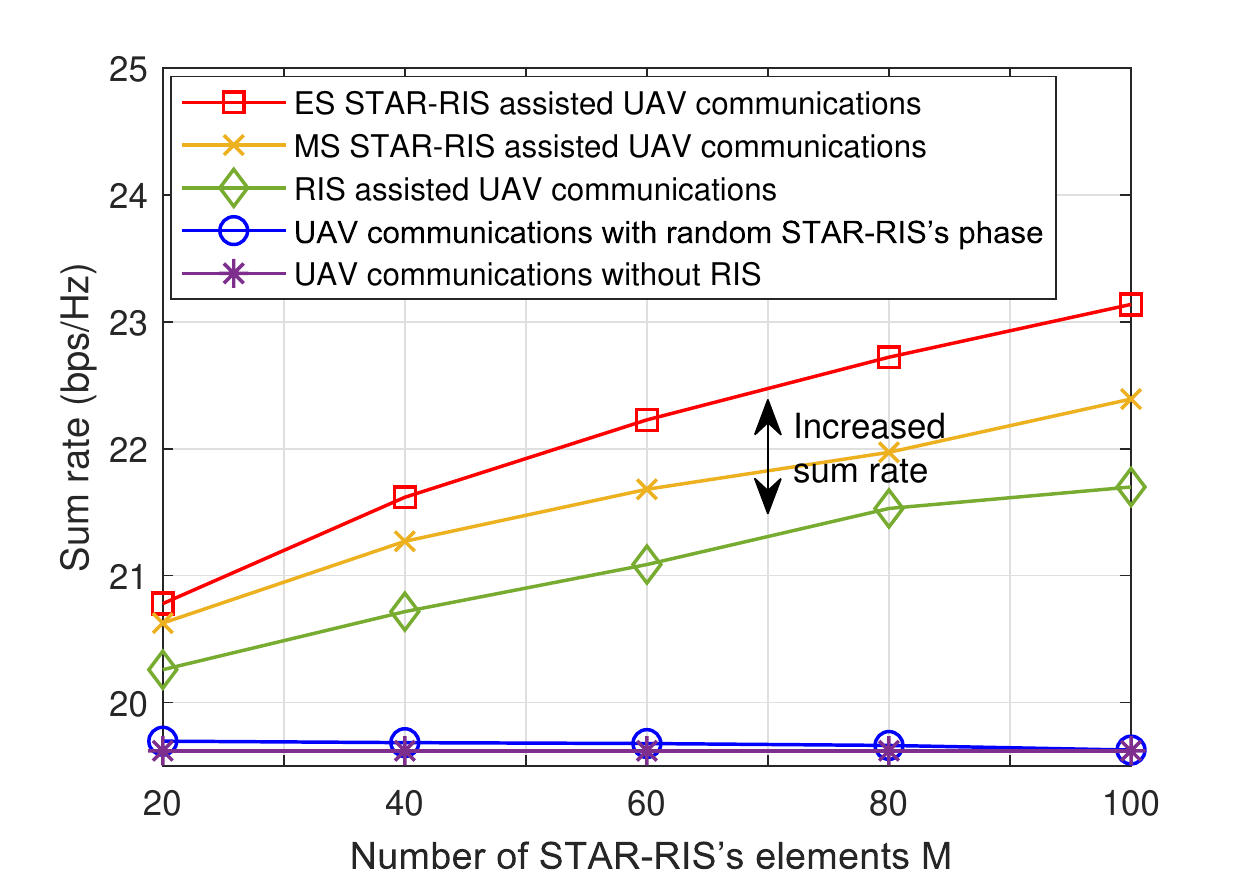}}\label{Sum_rate_m}
\subfigure[]{
\includegraphics[width=2.32in]{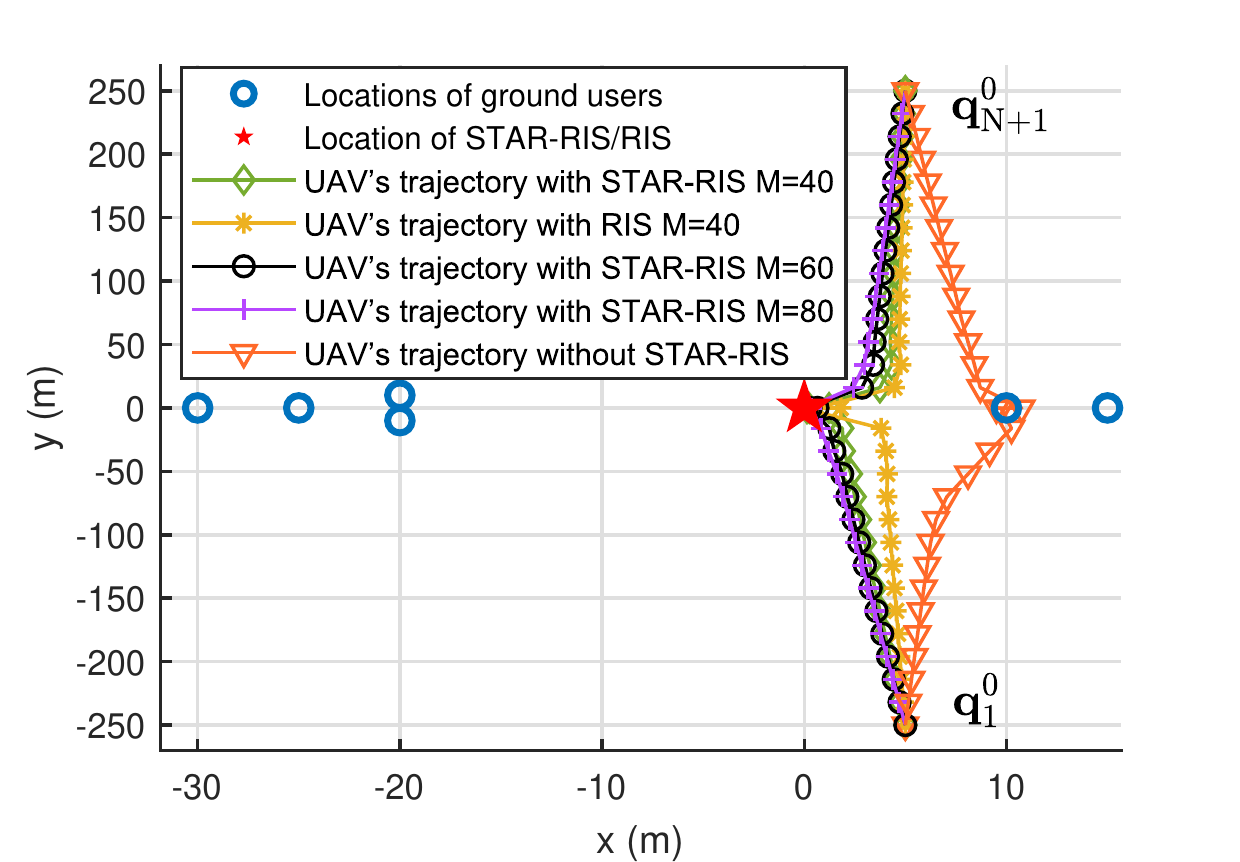}}\label{UAV_Q}
\subfigure[]{
\includegraphics[width=2.32in]{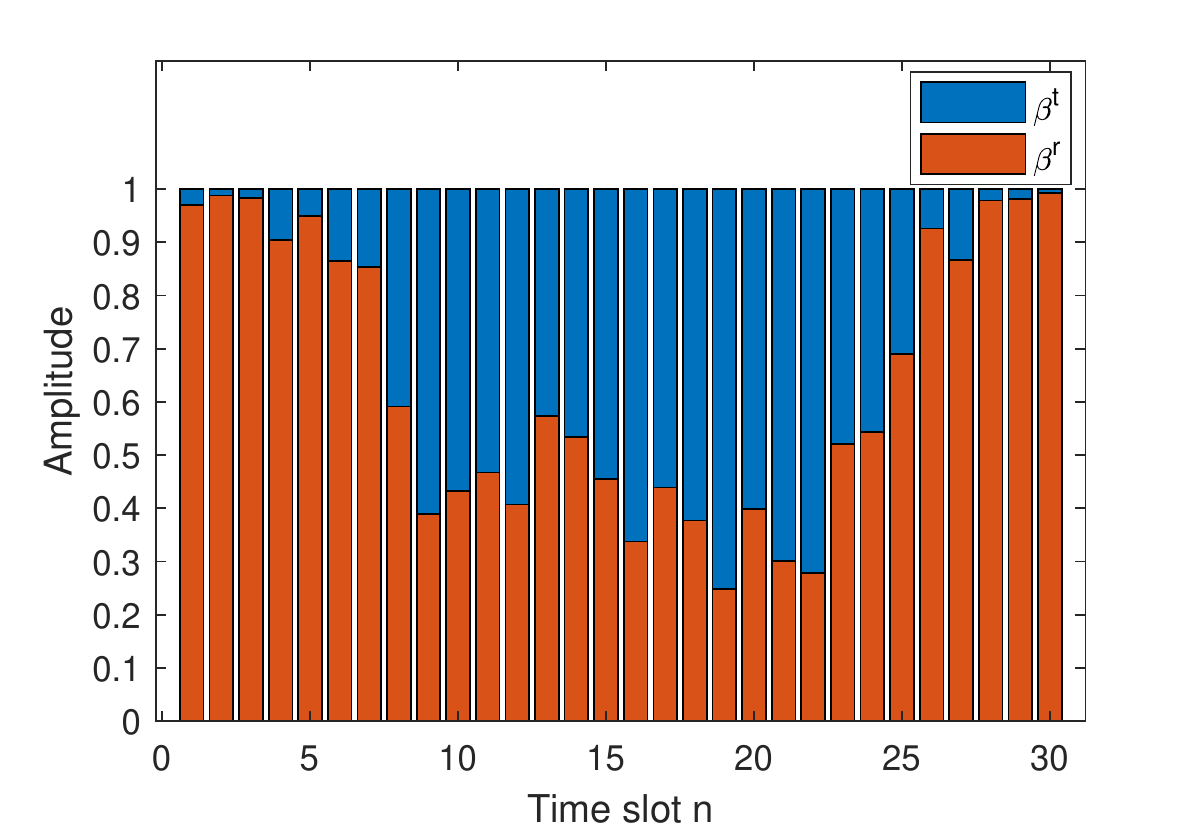}}\label{beta_n}
\vspace{-0.15in}
\caption{(a) Sum rate \emph{vs.} the number of STAR-RIS's elements, $N=30$. (b) The UAV's trajectories in different situations, $N=30$. (c) The average amplitudes of the transmission and reflection coefficients of all elements \emph{vs.} time slot $n$, $M=40$.}
\vspace{-0.15in}
\end{figure*}

In Fig. 2(a), the sum rate of users versus the number of STAR-RIS's elements is illustrated. It is found that with the assistance of STAR-RIS or RIS, the sum rate of users grows as the number of elements increases since more scattering elements bring higher beamforming gains at the users. More importantly, the STAR-RIS with ES outperforms the STAR-RIS with MS or traditional RIS. This is because the STAR-RIS for ES owns extra DoFs by splitting the energy of reflection and transmission for each element, while each element for MS can only operate in a binary mode, i.e., either in the reflection or in the transmission mode. The traditional RIS exhibits much worse performance as it can only reflect or transmit signals by a fixed number of elements.

Fig. 2(b) illustrates the trajectory of the UAV, where the users are located at both sides of STAR-RIS. It can be observed that the UAV always flies and serves users at the right side of STAR-RIS, i.e., a single STAR-RIS can achieve full-space coverage, while it is impossible for a transmitting/reflecting-only RIS. Besides, the UAV's trajectory with STAR-RIS is closer to the location of STAR-RIS compared to that with RIS. This is because the STAR-RIS can provide extra DoFs to further improve channel conditions by flexible reflecting/transmitting energy splitting. Such an observation reveals the potential advantages of STAR-RIS again.

Fig. 2(c) presents the average amplitudes of the reflection and transmission coefficients of all elements in the STAR-RIS at different time slot. It can be seen that at the first and last several time slots, i.e., when the UAV is far away from STAR-RIS, the amplitudes of reflection coefficients are larger than that of transmission coefficients. This is because the long distance leads to poor channel conditions between UAV and STAR-RIS. Thus, more energy is allocated to the reflected signals to acquire higher sum rate as the users in reflection space are closer to the UAV and STAR-RIS. On the contrary, when the UAV is closer to STAR-RIS at the middle couple of time slots, better channel conditions between UAV and STAR-RIS can be obtained. Since there are more users in the transmission space, the STAR-RIS allocates more energy to the transmission mode in order to make better use of the passive beamforming gain and improve the sum rate of users.

\vspace{-0.05in}
\section{Conclusion}
In this paper, we studied the novel STAR-RIS assisted UAV communications, where the sum rate of all users was maximized by decomposing the formulated problem into three subproblems. Simulations showed that the STAR-RIS can achieve higher sum rate than traditional RIS owing to the extra DoFs. Moreover, the UAV's trajectory was closer to STAR-RIS, and the energy splitting for reflection and transmission highly depended on the real-time trajectory of UAV.

\vspace{-0.1in}
\bibliographystyle{IEEEtran}
\bibliography{IEEEabrv,bib2014}

\begin{thebibliography}{10}
\providecommand{\url}[1]{#1}
\csname url@samestyle\endcsname
\providecommand{\newblock}{\relax}
\providecommand{\bibinfo}[2]{#2}
\providecommand{\BIBentrySTDinterwordspacing}{\spaceskip=0pt\relax}
\providecommand{\BIBentryALTinterwordstretchfactor}{4}
\providecommand{\BIBentryALTinterwordspacing}{\spaceskip=\fontdimen2\font plus
\BIBentryALTinterwordstretchfactor\fontdimen3\font minus
  \fontdimen4\font\relax}
\providecommand{\BIBforeignlanguage}[2]{{%
\expandafter\ifx\csname l@#1\endcsname\relax
\typeout{** WARNING: IEEEtran.bst: No hyphenation pattern has been}%
\typeout{** loaded for the language `#1'. Using the pattern for}%
\typeout{** the default language instead.}%
\else
\language=\csname l@#1\endcsname
\fi
#2}}
\providecommand{\BIBdecl}{\relax}
\BIBdecl

\bibitem{wu2019towards}
Q.~Wu and R.~Zhang, ``Towards smart and reconfigurable environment: Intelligent
  reflecting surface aided wireless network,'' \emph{IEEE Communications
  Magazine}, vol.~58, no.~1, pp. 106--112, Jan. 2020.

\bibitem{niu2021simultaneous}
H.~Niu \emph{et~al.}, ``Simultaneous transmission and reflection reconfigurable
  intelligent surface assisted secrecy {MISO} networks,'' \emph{IEEE
  Communications Letters}, vol.~25, no.~11, pp. 3498--3502, Nov. 2021.

\bibitem{niu2022weighted}
H.~Niu, Z.~Chu, F.~Zhou, P.~Xiao, and N.~Al-Dhahir, ``Weighted sum rate
  optimization for {STAR-RIS}-assisted {MIMO} system,'' \emph{IEEE Transactions
  on Vehicular Technology}, vol.~71, no.~2, pp. 2122--2127, Feb. 2022.

\bibitem{liu2021star}
Y.~Liu, X.~Mu, J.~Xu \emph{et~al.}, ``{STAR}: Simultaneous transmission and
  reflection for 360{$^{\circ}$} coverage by intelligent surfaces,'' \emph{IEEE
  Wireless Communications}, vol.~28, no.~6, pp. 102--109, Dec. 2021.

\bibitem{zuo2021joint}
J.~Zuo, Y.~Liu, Z.~Ding \emph{et~al.}, ``Joint design for simultaneously
  transmitting and reflecting ({STAR}) {RIS} assisted {NOMA} systems,''
  \emph{IEEE Transactions on Wireless Communications},
  10.1109/TWC.2022.3197079, 2022.

\bibitem{mu2021simultaneously}
X.~Mu, Y.~Liu, L.~Guo \emph{et~al.}, ``Simultaneously transmitting and
  reflecting ({STAR}) {RIS} aided wireless communications,'' \emph{IEEE
  Transactions on Wireless Communications}, vol.~21, no.~5, pp. 3083--3098,
  May. 2022.

\bibitem{niu2022weighted2}
H.~Niu and X.~Liang, ``Weighted sum-rate maximization for {STAR}-{RIS}s-aided
  networks with coupled phase-shifters,'' \emph{IEEE Systems Journal},
  10.1109/JSYST.2022.3159551, 2022.

\bibitem{liu2021simultaneously}
Y.~Liu, X.~Mu, R.~Schober, and H.~V. Poor, ``Simultaneously transmitting and
  reflecting ({STAR})-{RIS}s: A coupled phase-shift model,'' in \emph{Proc.
  IEEE ICC}, 2022, pp. 2840--2845.

\bibitem{li2020reconfigurable}
S.~Li, B.~Duo \emph{et~al.}, ``Reconfigurable intelligent surface assisted
  {UAV} communication: Joint trajectory design and passive beamforming,''
  \emph{IEEE Wireless Communications Letters}, vol.~9, no.~5, pp. 716--720,
  May. 2020.

\bibitem{hua2021uav}
M.~Hua, L.~Yang, Q.~Wu, C.~Pan \emph{et~al.}, ``{UAV}-assisted intelligent
  reflecting surface symbiotic radio system,'' \emph{IEEE Transactions on
  Wireless Communications}, vol.~20, no.~9, pp. 5769--5785, Sept. 2021.

\bibitem{li2021robust}
S.~Li, B.~Duo, M.~Di~Renzo \emph{et~al.}, ``Robust secure {UAV} communications
  with the aid of reconfigurable intelligent surfaces,'' \emph{IEEE
  Transactions on Wireless Communications}, vol.~20, no.~10, pp. 6402--6417,
  Oct. 2021.

\bibitem{mu2020exploiting}
X.~Mu, Y.~Liu, L.~Guo \emph{et~al.}, ``Exploiting intelligent reflecting
  surfaces in {NOMA} networks: Joint beamforming optimization,'' \emph{IEEE
  Transactions on Wireless Communications}, vol.~19, no.~10, pp. 6884--6898,
  Oct. 2020.

\end{thebibliography}

\end{document}